\documentclass[showpacs]{revtex4}
\usepackage{epsfig}
\def\beq{\begin{equation}}
\def\eeq{\end{equation}}
\def\beqa{\begin{eqnarray}}
\def\eeqa{\end{eqnarray}}

\def\GeV{\nobreak\,\mbox{GeV}}

\def\pli{p^\prime}
\def\ql{{p^\prime}^2}

\def\muli{\mu^\prime}
\def\nuli{\nu^\prime}
\def\ali{\alpha^\prime}

\begin{document}
\title{$J/\psi D^*D^*$ vertex from QCD sum rules.}
\author{M.E. Bracco, M. Chiapparini}
\affiliation{Instituto de F\'{\i}sica, Universidade do Estado do Rio de 
Janeiro, 
Rua S\~ao Francisco Xavier 524, 20550-900 Rio de Janeiro, RJ, Brazil}
\author{F.S. Navarra and M. Nielsen}
\affiliation{Instituto de F\'{\i}sica, Universidade de S\~{a}o Paulo, 
C.P. 66318, 05389-970 S\~{a}o Paulo, SP, Brazil.}

\begin{abstract}
We calculated the strong form factor and coupling constant for the 
$J/\psi D^* D^*$ vertex in a QCD sum rule calculation. We performed a 
double Borel sum rule for the three point correlation function of vertex
 considering both $J/\psi$ and $D^*$ mesons 
off--shell. The form factors obtained are very different, but they give 
the same coupling constant. 
\end{abstract}


\maketitle


An important topic of RHIC physics is charmonium production \cite{kha}. The
first results have already been reported \cite{phenix}, but they did not reach
yet the quality of the previous lower energies measurements performed at
CERN-SPS, where an anomalous suppression was found in the most
central collisions \cite{NA50}.

Understanding charmonium production and destruction in these highly complicated
collisions is a challenge for theorists, and in the last six years an intense
effort was dedicated to understand charmonium interactions with light hadrons.
This subject can be divided into high ($\sqrt{s}\geq10\GeV$) and low
($\sqrt{s}\leq10\GeV$) energy interactions. The latter have been treated mostly
with the help of effective Lagrangians. In this approach one has a better control 
of the relevant symmetries, which dictate the dynamics 
\cite{mamu,linko,haglin,su,regina}. However, very soon it was realized that
working with effective lagrangians requires a very good knowledge of the form
factors associated with vertices involving charmed mesons as, for example,
the $D^*D\pi$ vertex, to mention the most famous one. Choosing a softer or harder
form factor may change the final cross section up to two orders of magnitude. This
dramatic change outshines the detailed discussion concerning the role played by 
gauge invariance, anomalous parity couplings and other dynamical features of the
interaction.

Until four years ago nothing was known about such form factors. At that time we 
launched a program of calculating these important quantities in the framework
of QCD sum rules \cite{svz}. Since then we have been continuously attacking this 
problem and computing different vertices 
\cite{ddpi,ddpir,ddrho,matheus1,matheus2,dasilva,Angelo}. In doing so, we have also
improved the strategies to determine the coupling constants. One of them is
experimentaly measured \cite{ddpiex} but the others have to be estimated
using a vector meson dominance analysis of some measured decays, or using SU(4)
symmetry relations. In some cases our results gave support to these more
empyrical estimates and in some other they did not.

As it frequently happens, during the execution of the project the original 
motivation was extended to a wider scope of questions concerning charmed mesons.
A particularly interesting side product of our works, \cite{ddrho,matheus2} and
also the present one, is the conclusion regarding the ``resolving power''
of a compact $(J/\psi)$ or a extended probe $(\rho)$ hitting the $D$ meson.
It was reassuring to observe how the $J/\psi$ behaves more like a point-like
parton penetrating the $D$ meson, whereas the $\rho$ behaves more like a large
hadron being able to measure the size of the $D$ meson. All this information is
encoded in the form factors. In what follows we will have a glimpse on how the
$J/\psi$ ``sees'' a $D^*$  meson. From spin symmetry of
the heavy quark effective theory (HQET) one would expect a similar behaviour as
observed in the $J/\psi DD$ vertex \cite{matheus2}. 

The present calculation is part of this project. Although the method is the same
as used before, because there are three vector particles involved, the
number of Lorentz structures is much bigger and the calculation is considerably
more involved. On the other hand, since the three particles are heavy, we feel
more confident about neglecting some higher dimension contributions to the
operator product expansion (OPE). We will compare our results with the ones from 
other models \cite{deandrea,haglin2} and we will also check if the use of HQET 
symmetry is appropriate.

Following the QCDSR formalism described in our previous works 
\cite{ddpi,ddpir,ddrho}, we write the three-point function associated 
with the $J/\psi D^*D^*$ vertex, which is given by
\begin{equation}
\Gamma_{\nu \alpha \mu}^{(D^*)}(p,\pli)=\int d^4x \, d^4y \;\;
e^{i\pli\cdot x} \, e^{-i(\pli-p)\cdot y}
\langle 0|T\{j_{\nu}^{D^*}(x) j_{\alpha}^{\bar {D^*}}(y) 
 j_{\mu}^{\psi \dagger}(0)\}|0\rangle\, \label{cordoff} 
\end{equation}
for an off-shell $D^* $ meson, and:
\begin{equation}
\Gamma_{\nu \alpha \mu}^{(J/\psi)}(p,\pli)=\int d^4x \, 
d^4y \;\; e^{i\pli\cdot x} \, e^{-i(\pli-p)\cdot y}\;
\langle 0|T\{j_{\nu}^{D^*}(x)  j_{\mu}^{\psi \dagger}(y) 
 j_{\alpha}^{D^* \dagger}(0)\}|0\rangle\, ,\label{corjpsioff} 
\end{equation}
for an off-shell $J/\psi$ meson. The general expression for the vertices 
(\ref{cordoff}) and (\ref{corjpsioff}) 
has fourteen independent Lorentz structures. We can write 
$\Gamma_{\nu \alpha \mu}$ in terms of the invariant amplitudes associated 
with each one of these structures in the following form:
\begin{eqnarray}
\Gamma_{\mu\nu\alpha}(p,\pli)&=&
    \Gamma_1(p^2 , \ql , q^2) g_{\mu \nu} p_{\alpha} 
  + \Gamma_2(p^2,\ql, q^2) g_{\mu \alpha} p_{\nu} 
  + \Gamma_3(p^2,\ql , q^2) g_{\nu \alpha} p_{\mu} 
  + \Gamma_4(p^2,\ql ,q^2) g_{\mu \nu} \pli_{\alpha} \nonumber \\ 
&&+ \Gamma_5(p^2, \ql ,q^2) g_{\mu \alpha} \pli_{\nu} 
  + \Gamma_6(p^2,\ql ,q^2) g_{\nu\alpha} \pli_{\mu}  
  + \Gamma_7(p^2,\ql ,q^2) p_{\mu} p_{\nu} p_{\alpha}
  + \Gamma_8(p^2,\ql ,q^2) \pli_{\mu} \pli_{\nu} p_{\alpha}  \nonumber \\
&&+ \Gamma_9(p^2,\ql ,q^2) p_{\mu} \pli_{\nu} p_{\alpha}  
  + \Gamma_{10}(p^2,\ql ,q^2) p_{\mu} p_{\nu} \pli_{\alpha} 
  + \Gamma_{11}(p^2,\ql ,q^2) \pli_{\mu} \pli_{\nu} p_{\alpha}
  + \Gamma_{12}(p^2,\ql ,q^2) \pli_{\mu} p_{\nu} \pli_{\alpha}  \nonumber \\
&&+ \Gamma_{13}(p^2,\ql ,q^2) p_{\mu} \pli_{\nu} \pli_{\alpha} 
  + \Gamma_{14}(p^2,\ql ,q^2) \pli_{\mu} \pli_{\nu} \pli_{\alpha}
  \label{trace}  
\end{eqnarray}

The correlator function, Eqs.~(\ref{cordoff}) and 
(\ref{corjpsioff}), can be calculated in two diferent ways: using quark 
degrees of freedom --the theoretical or QCD side-- or using hadronic 
degrees of freedom --the phenomenological side. In the QCD side the 
correlators 
is evaluated by using the 
Wilson operator product expansion (OPE). The OPE incorporates the effects 
of the QCD vacuum through an infinite serie of condensates of in\-crea\-sing 
dimension. On the other hand, the representation in terms of 
hadronic degrees of freedom is responsable for the introduction of the form 
factors, decay constants and masses. Both representations are matched using 
the quark-hadron global duality.

In the QCD side, each meson interpolating
field appearing in Eqs.~(\ref{cordoff}) and (\ref{corjpsioff}) can be written 
in terms of the quark field operators in the following form: 
$j_{\nu}^{D^*}(x) = \bar c(x) \gamma_{\nu} q(x)$ and
$j_{\mu}^{\psi}(x) = \bar c(x) \gamma_{\mu} c(x)$, where 
$q$ and $c$ are the up/down and charm quark field respectively. Each 
one of these currents have the same quantum numbers as the associated
meson. 
 
\begin{figure}[h]
\begin{picture}(12,3.5)
\put(0.0,0.5){\vector(1,0){1.5}}
\put(3.5,0.5){\vector(-1,0){2}}
\put(3.5,0.5){\vector(1,0){1.5}}
\put(1.5,0.5){\vector(1,1){1}}
\put(2.5,1.5){\vector(1,-1){1}}
\put(2.5,1.5){\vector(0,1){1.5}}
\put(2.65,2.75){$q_\alpha$}
\put(0.25,0.65){$p_\mu$}
\put(4.55,0.65){$p'_\nu$}
\put(2.4,0.2){$\bar c$}
\put(1.85,1.1){$c$}
\put(3,1.1){$q$}
\put(2.4,1.2){$y$}
\put(1.75,0.53){$0$}
\put(3.05,0.53){$x$}
\put(1.95,2.2){$D^*$}
\put(0.35,0.1){$J/\psi$}
\put(4,0.1){$\bar{D^*}$}
\put(7,0.5){\vector(1,0){1.5}}
\put(10.5,0.5){\vector(-1,0){2}}
\put(10.5,0.5){\vector(1,0){1.5}}
\put(8.5,0.5){\vector(1,1){1}}
\put(9.5,1.5){\vector(1,-1){1}}
\put(9.5,3){\vector(0,-1){1.5}}
\put(9.65,2.75){$q_\alpha$}
\put(7.25,0.65){$p_\mu$}
\put(11.55,0.65){$p'_\nu$}
\put(9.4,0.2){$\bar q$}
\put(8.85,1.1){$c$}
\put(10,1.1){$c$}
\put(9.4,1.2){$y$}
\put(8.75,0.53){$0$}
\put(10.05,0.53){$x$}
\put(8.75,2.2){$J/\psi$}
\put(7.35,0.1){$D^*$}
\put(11,0.1){$D^*$}
\end{picture}
\caption{Perturbative diagrams for $D^*$ off shell (left) and $J/\psi$
off shell (right).}
\label{fig1}
\end{figure}
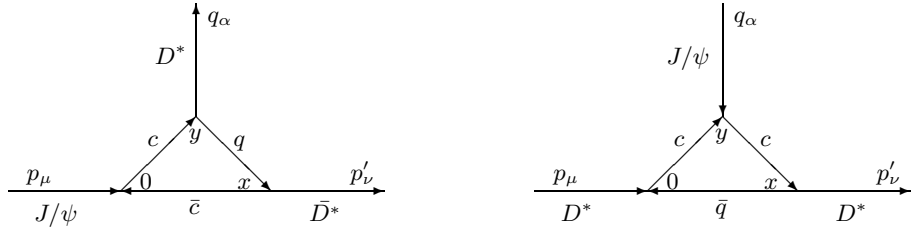

For each one of the invariant amplitudes appearing in Eq.(\ref{trace}), we 
can write a double dispersion relation over the virtualities $p^2$ and 
${\pli}^2$, holding $Q^2= -q^2$ fixed:
\begin{equation}
\Gamma_i(p^2,{\pli}^2,Q^2)=-\frac{1}{\pi^2}\int_{s_{min}}^\infty ds
\int_{m_c^2}^\infty du \:\frac{\rho_i(s,u,Q^2)}{(s-p^2)(u-{\pli}^2)}\;,
\;\;\;\;\;\;i=1,\ldots,14 \label{dis}
\end{equation}
where $\rho_i(s,u,Q^2)$ equals the double discontinuity of the amplitude
$\Gamma_i(p^2,{\pli}^2,Q^2)$, calculated using the Cutkosky's rule, and
where $s_{min}=4m_c^2$ in the case of $D^*$ off-shell, and
$s_{min}=m_c^2$ in the case of $J/\psi$ off-shell, with $m_c$ being the mass
of the charm quark. The invariant amplitudes receive contributions 
from all terms in the OPE. The first one of those contributions comes from 
the perturbative term and it is represented in Fig.~\ref{fig1}.

In principle, we can work with any structure appearing in
Eq.(\ref{trace}). But we must choose those which have less ambiguities in 
the QCD sum rules approach, which means, less influence of the condensates of 
higher dimension, and a better stability as a function of the Borel mass.
A possibility could be work with the structures 
that present some  symmetry in the momenta of the mesons, for 
example, $p_{\mu} p_{\nu} p_{\alpha}$ or 
$\pli_{\mu} \pli_{\nu} \pli_{\alpha}$. In these cases, however, the quark 
condensate is absent. Since the quark condensate is the condensate of lower 
dimension we will choose structures in which the quark condensate contribute. 
These structures are 
$g_{\mu\nu} (\pli +p)_\alpha$, $g_{\alpha\nu}(\pli - p )_\mu$ and 
$g_{\alpha\mu}(\pli - p)_\nu$. 

In this paper we use the structure $g_{\mu\nu} (\pli + p )_\alpha$,
which presented a better behavior. The corresponding perturbative spectral 
densities which enters in Eq.~(\ref{dis}) is 
\begin{equation}
\rho^{(D^*)}(s,u,Q^2)=\frac{3}{2\pi\sqrt\lambda}
\left[\left(\frac{u+s-t}{2}\right)\left(\frac{M+N}{2}\right) 
+ 2(C+D)+\frac{\pi}{2}(3m^2-u-s) - 2 G \right] \label{sdjpof1}
\end{equation}
for $D^*$ off-shell, and
\begin{equation}
\rho^{(J/\psi)}(s,u,Q^2)=-\frac{3}{2\pi\sqrt\lambda} 
\left[\left(m^2+\frac{t-u-s}{2}\right)
\left(\frac{M+N}{2}\right) - 2 (C+D)
+\frac{\pi}{2}(u+s-2 m^2) + 2 G\right] 
\label{sddof2}
\end{equation}
for $J/\psi$ off-shell. Here $\lambda = \lambda(s,u,t) = 
s^2+t^2+u^2-2st-2su-2tu$, $s=p^2$, $u=p'^2$, $t=-Q^2$ and $C$, $D$, 
$G$, $M$ and $N$ are functions of $(s,t,u)$, given by 
the following expressions:
\begin{eqnarray}
C&=&\frac{\pi{\overline{|\vec k|}}^2}{\sqrt{s}}
\left(1 - \cos^2\overline{\theta}\right)
\left(\frac{\overline{|\vec k|} p'_0}{|\vec p'|}
\cos \overline{\theta}-\overline{k_0}\right)
\label{C} \\
D&=&-\frac{\pi\overline{|\vec k|}^3 }{|\vec p'|} 
\left(1 - \cos^2\overline{\theta}\right) \cos\overline{\theta}\label{D} \\
G&=&-\pi \overline{|\vec k|}^2 
\left(1 - \cos^2\overline{\theta} \right) \label{G} \\
M&=&\frac{2\pi}{\sqrt{s}}\left( \overline{k_0}
- \frac{\overline{|\vec k|}p'_0}{|\vec p'|} 
\cos\overline{\theta}\right) \label{M} \\
N&=&2\pi\frac{\overline{|\vec k|}}{ |\vec p'|} \cos\overline{\theta}
\label{N} \\
p'_0&=&\frac{s+u-t}{2\sqrt{s}}    \label{pl0} \\
|\vec p'|^2&=&\frac{\lambda}{4s}     \label{vpl}\\
\overline{k_0}&=& \frac{s-m^2}{2\sqrt{s}}   \label{k0b} \\
\overline{|\vec k|}^2&=& \overline{k_0}^2-m^2   \label{vk} \\
\cos\overline{\theta}&=-&\frac{u+\eta m_c^2-2p'_0\overline{k_0}}
{2|\vec p'|\overline{|\vec k|}}   \label{ctheta} 
\end{eqnarray}
where $m=0$ and $\eta=1$ for $D^*$ off-shell and $m=m_c$ and $\eta=-1$ 
for $J/\psi$ off-shell.

\begin{figure}[h]
\begin{picture}(6,3.5)
\put(0,0.5){\vector(1,0){1.5}}
\put(1.5,0.5){\line(1,0){0.75}}
\put(2.75,0.5){\line(1,0){0.75}}
\put(2.25,0.5){\circle*{0.15}}
\put(2.75,0.5){\circle*{0.15}}
\put(3.5,0.5){\vector(1,0){1.5}}
\put(1.5,0.5){\vector(1,1){1}}
\put(2.5,1.5){\vector(1,-1){1}}
\put(2.5,3){\vector(0,-1){1.5}}
\put(2.65,2.75){$q_\alpha$}
\put(0.25,0.65){$p_\mu$}
\put(4.55,0.65){$p'_\nu$}
\put(2.225,0.17){$\langle \bar q q\rangle$}
\put(1.85,1.1){$c$}
\put(3,1.1){$c$}
\put(2.4,1.2){$y$}
\put(1.75,0.53){$0$}
\put(3.05,0.53){$x$}
\put(1.75,2.2){$J/\psi$}
\put(0.35,0.1){$D^*$}
\put(4,0.1){$D^*$}
\end{picture}
\caption{Contribution of the $q\bar q$ condensate for $J/\psi$ off-shell
case.}
\label{fig2}
\end{figure}
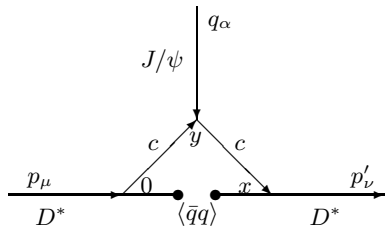

The contribution of the quark condensate which survives after the
double Borel transform is represented in Fig.~\ref{fig2} for the 
$J/\psi$ off-shell case, and is given by 
\begin{equation}
\Gamma_c^{(J/\psi)} = -\frac{m_c \langle \bar q q\rangle}
{(p^2 - m_c^2)(p'^2 -m_c^2)} 
\end{equation}
where $\langle \bar q q\rangle$ is the light quark condensate. For the
$D^*$ off-shell there is a similar contribution, this time
proportional to the mass of the light quark and to the $\bar c c$
condensate, both small, therefore, we will not consider it here. 
We also expect that the perturbative contribution be the dominant one in the 
OPE, because we are dealing with heavy quarks. For this reason, we do not 
include the gluon and quark-gluon condesates in the present work.

The resulting vertex functions in the QCD side for the structure 
$g_{\mu\nu} (\pli + p )_\alpha$ are written as
\begin{equation}
\Gamma^{(D^*)}(p,\pli)=
-\frac{1}{4\pi^2}\int_{4m_c^2}^{s_0} ds
\int_{m_c^2}^{u_0} du \:
\frac{\rho^{(D^*)}(s,u,Q^2)}{(s-p^2)(u-{\pli}^2)} \label{qcdd}
\end{equation}
for $D^*$ off-shell and
\begin{equation}
\Gamma^{(J/\psi)}(p,\pli)=
-\frac{1}{4\pi^2}\int_{m_c^2}^{s_0} ds
\int_{m_c^2}^{u_0} du \:
\frac{\rho^{(J/\psi)}(s,u,Q^2)}{(s-p^2)(u-{\pli}^2)}+
\Gamma_c^{(J/\psi)} \label{qcdjp}
\end{equation}
for $J/\psi$ off-shell, where, as usual, we have already transfered the 
continuum 
contribution from the hadronic side to the QCD side, through the introduction
of the continuum thresholds $s_0$ and $u_0$ \cite{io2}.

The phenomenological side of the vertex function is
obtained by considering the contribution of the $J/\psi $ and two $D^*$ 
mesons in Eqs.~(\ref{cordoff}) and (\ref{corjpsioff}).  The effective
Lagrangian for the $J/\psi D^*D^*$ interaction is given by \cite{linko}
\beqa
&&\mathcal{L}_{J/\psi D^{*}D^{*}}=
ig_{\psi D^{*}D^{*}} \Big [\psi^{\mu} \Big ((\partial_{\mu}D^{*\nu})
\bar{D^{*}_{\nu}}
-D^{*\nu}\partial_{\mu}\bar{D^{*}_{\nu}} \Big )
+ \Big ((\partial_{\mu}\psi^{\nu})D^{*}_{\nu}
\nonumber\\
&&-\psi^{\nu}\partial_{\mu}D^{*}_{\nu} \Big)\bar{D^{*\mu}}
+D^{*\mu}\Big(\psi^{\nu}\partial_{\mu}\bar{D^{*}_{\nu}}
-(\partial_{\mu}\psi^{\nu})\bar{D^{*}_{\nu}} \Big) \Big],
\eeqa
from where one can extract the matrix element associated with the
$J/\psi D^*D^*$ vertex:
\beqa
&&\langle D^*(p_1,\alpha)D^*(p_2,\nu)|J/\psi(p_3,\mu)\rangle=
\nonumber\\
&-&i g_{J/\psi D^*D^*}[(p_2-p_1)_\mu g_{\alpha\nu}-(p_3+p_2)_\mu g_{\mu\nu}
+(p_3+p_1)_\mu g_{\mu\alpha}]\epsilon^\mu_{J/\psi}(p_3)\epsilon{^{\alpha*}}_{D^*}
(p_1)\epsilon^{\nu*}_{D^*}(p_2).
\label{coup}
\eeqa
The meson decay constants, $f_{D^*}$ and $f_{J/\psi}$, are
defined by the matrix elements
\beq
\langle 0|j^{\mu}_{D^*}|{D^*(p)}\rangle= m_{D^*} f_{D^*} \epsilon^{\mu*}_{D^*}
(p)
\label{fd}
\eeq
and
\beq
\langle 0|j_{\nu}^{J/\psi}|{J/\psi(p)}\rangle= m_{J/\psi} f_{J/\psi} 
\epsilon^{\nu*}_{J/\psi}(p) \label{fpsi}
\eeq
where $\epsilon^{\mu}_{D^*}$ and $\epsilon^{\nu}_{J/\psi}$ are the
polarization vectors of the mesons $D^*$ and $J/\psi$ respectively. 
Saturating Eqs.~(\ref{cordoff}) and (\ref{corjpsioff}) with the
$J/\psi $ and two $D^*$ states and using Eqs.~(\ref{coup}), (\ref{fd})
and (\ref{fpsi}) we arrive at
\beqa
&&\Gamma_{\nu\alpha\mu}^{(J/\psi)}= -g^{(J/\psi)}_{J/\psi D^*D^*}(Q^2)
\frac{f^2_{D^*}f_{J/\psi}m^2_{D^*}m_{J/\psi}}
{(P^2+m^2_{D^*})(Q^2+m^2_{J/\psi})({P^\prime}^2 +m^2_{D^*})}
\left(-g_{\mu\muli}-{p_\mu p_{\muli}\over m^2_{D^*}}\right)\times\nonumber\\
&&\left(-g_{\nu\nuli}-{\pli_\nu \pli_{\nuli}\over m^2_{D^*}}\right)
\left(-g_{\alpha\ali}-{q_\alpha q_{\ali}\over m^2_{J/\psi}}\right)
\left[(p+\pli)^{\ali}g^{\muli\nuli}+(q-p)^{\nuli}g^{\ali\muli}-
(\pli-q)^{\muli}g^{\ali\nuli}\right],\label{allstru}
\eeqa
when the $J/\psi$ is off-shell, with a similar expression for the
$D^*$ off-shell. The contractions of $\muli,~\nuli$ and $\ali$ in the above 
equation will lead to the fourteen Lorentz structures appearing in 
Eq.~(\ref{trace}).
We can also see from Eq.~(\ref{allstru}) that the form factor
$g^{(J/\psi)}_{J/\psi D^*D^*}(Q^2)$ is the same for all the structures
and thus can be extracted from sum rules written for any of these
structures.

The 
resulting phenomenological invariant amplitudes associated with the structure 
$g_{\mu\nu} (\pli + p )_{\alpha}$ are 
\begin{equation}
\Gamma^{(D^*)}_{ph}(p^2,{\pli}^2,Q^2)= g^{(D^*)}_{J/\psi D^*D^*}(Q^2)
\frac{f_{D^*}^2f_{J/\psi}m^2_{D^*}m_{J/\psi}}
{(P^2+m^2_{J/\psi})(Q^2+m^2_{D^*})(P^{\prime2} +m^2_{D^*})}
\label{phendsoff1}
\end{equation}
for the $D^*$ off-shell, and 
\begin{equation}
\Gamma^{(J/\psi)}_{ph}(p^2,{\pli}^2,Q^2)= g^{(J/\psi)}_{J/\psi D^*D^*}(Q^2)
\frac{f^2_{D^*}f_{J/\psi}m^2_{D^*}m_{J/\psi}}
{(P^2+m^2_{D^*})(Q^2+m^2_{J/\psi})(P^{\prime2} +m^2_{D^*})}
\label{phenjpoff}
\end{equation}
for $J/\psi$ off-shell.

To improve the matching between the two sides of the sum rules
we perform a double Borel transformation \cite{io2} in the variables 
$P^2=-p^2\rightarrow M^2$ and $P'^2=-{\pli}^2\rightarrow M'^2$, 
on both invariant amplitudes $\Gamma$ and $\Gamma_{ph}$. 
Equating the results we get the final expression for the sum rule 
which allow us to obtain the form factors 
$g^{(T)}_{J/\psi D^*D^*}(Q^2)$ appearing in 
Eqs.~(\ref{phendsoff1})--(\ref{phenjpoff}), where $T$ is $D^*$ or  
$J/\psi$. 
In this work we use the following relations between the Borel masses $M^2$ and 
$M'^2$: $\frac{M^2}{M'^2} = \frac{m^2_{J/\psi}}{m^2_{D^*}}$
for an off-shell $D^*$  and $M^2 = M'^2$ for an off-shell $J/\psi$.

\begin{table}[h]
\begin{tabular}{|c|c|c|c|c|c|} \hline
$m_c (\GeV)$ & $m_{D^*} (\GeV)$ & $m_{J/\psi} (\GeV)$ & $f_{D^*} (\GeV)$ & 
$f_{J/\psi} (\GeV)$ & 
$\langle \bar q q \rangle (\GeV)^3$  \\ \hline 
1.3&2.01&3.1&0.240&0.405&$(-0.23)^3$ \\ \hline
\end{tabular}
\caption{Parameters used in the calculation.}
\label{tableparam}
\end{table}


The values of the parameters used in the present calculation are 
presented in Table \ref{tableparam}. The continuum thresholds are given by 
$s_0=(m+ \Delta_s)^2$ and $u_0=(m+\Delta_u)^2$, where $m$ 
is the mass of the incoming meson. We
used the experimental value for the $J/\psi$ decay constant $f_{J/\psi}$, and 
took
$f_{D^*}$ from ref.~\cite{khod}.

\begin{figure}[ht] 
\begin{center}
\epsfig{file=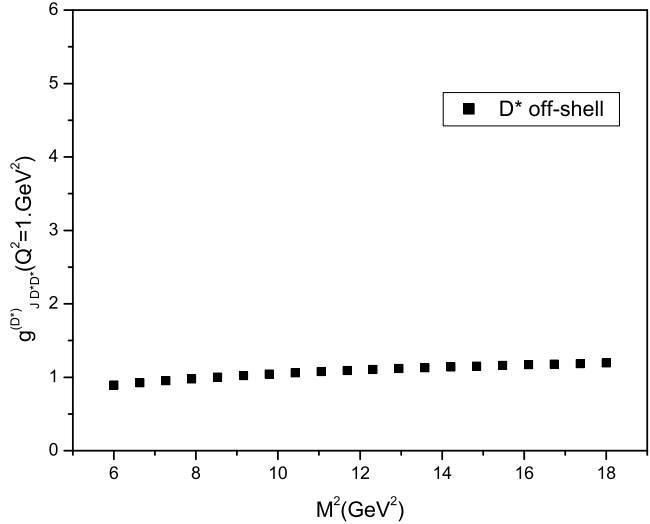}
\caption{Stability of $g^{(D^*)}_{J/\psi D^*D^*}(Q^2=1\GeV^2)$, as a function 
of the Borel mass $M^2$.}
\label{fig3}
\end{center}
\end{figure}
Using $\Delta_s=\Delta_u = 0.6 \GeV $ for the continuum thresholds 
and fixing $Q^2=1 \GeV^2$, we found a good stability of the 
sum rule for $g_{J/\psi D^* D^*}^{(D^*)}$ for $M^2$ in the interval 
$6< M^2 <18 \GeV^2$, as can be seen in Fig.~\ref{fig3}. 

\begin{figure}[ht] 
\begin{center}
\epsfig{file=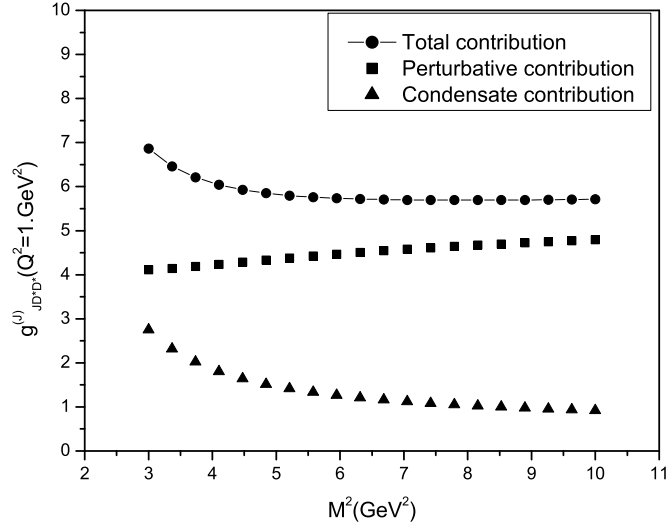}
\caption{Stability of $g^{(J/\psi)}_{J/\psi D^*D^*}(Q^2=1\GeV ^2)$, as a
function of the Borel mass. Here are represented the perturbative, condensate 
and total contributions.}
\label{fig4}
\end{center}
\end{figure}

In the case of 
$g_{J/ \psi D^* D^*}^{(J/\psi)}$ the interval for stability is 
$5<M^2 <10 \GeV^2 $, as can be seen in Fig.~\ref{fig4}. 

\begin{figure}[ht] 
\begin{center}
\epsfysize=4.0cm
\epsfig{file=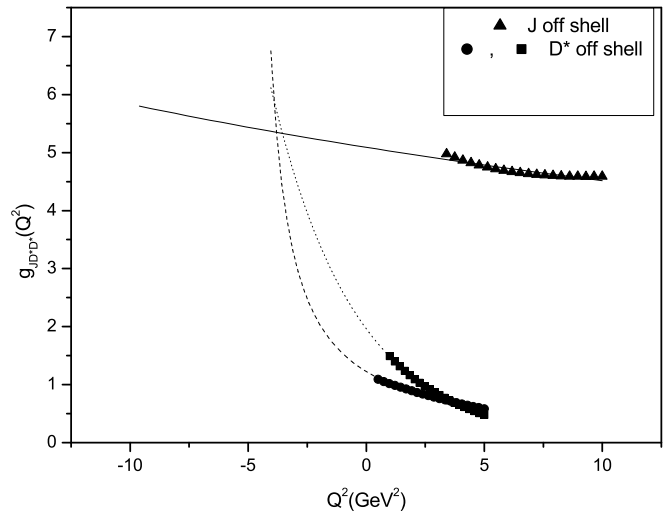}
\caption{$g^{(D^*)}_{J/\psi D^*D^*}$ (circles and squares) and 
$g^{(J/\psi)}_{J/\psi D^*D^*}$ (triangles) form factors as a function of
$Q^2$ from the QCDRS calculation of this work. The solid and dashed lines 
correspond to the  monopole parametrization of the QCDSR data, using the
two structures cited in the text, and the 
dotted line corresponds to the exponential parametrization.}
\label{fig5}
\end{center}
\end{figure}

Fixing 
$\Delta_s=\Delta_u=0.6\GeV$
we calculated the momentum dependence of the form factors. We present the 
results in Fig.~\ref{fig5}, where the circles corresponds to the 
$g_{J/ \psi D^* D^*}^{(D^*)}(Q^2)$ form factor (evaluated using $M^2=9.\GeV ^2$)
in the  interval where the 
sum rule is valid. The triangles are the result of the sum rule for the 
$g_{J/ \psi D^* D^*}^{(J/\psi)}(Q^2)$ form factor (evaluated using $M^2=6.\GeV ^2$). 

\vspace{0.5cm}

In the case of the off-shell $D^*$ meson, our numerical results can be 
fitted by a monopolar parametrization, dashed line in
Fig.~\ref{fig5}, in the following way:
\begin{equation}
g_{J/ \psi D^* D^*}^{(D^*)}(Q^2)= \frac{6.04}{Q^2+4.93}\;.
\label{monod}
\end{equation}
As in Ref.\cite{ddrho}, we define the coupling constant as the value of the 
form factor at $Q^2= -m^2_T$, where $m_T$ is the mass of the off-shell meson. 
Therefore, using $Q^2=-m_{D^*}^2$ in Eq~(\ref{monod}), the resulting coupling 
constant is:
\begin{equation}
g_{J/ \psi D^* D^*}= 6.75\;.  \label{gausscc}
\end{equation}

In the case of the off-shell $J/\psi$ meson, our sum rule result can also be 
fitted by a monopole parametrization, which is represented by the
solid line in Fig.~\ref{fig5}:  
\begin{equation}
g_{J/ \psi D^* D^*}^{(J/\psi)}(Q^2)=\frac{399.7}{Q^2+78.52}\;.
\label{monopole}
\end{equation}
Using $Q^2=-m_{J/\psi}^2$ in Eq~(\ref{monopole}) we get:
\begin{equation}
g_{J/ \psi D^* D^*}= 5.80,
\label{monopolecc}
\end{equation}
in a good agreement with the result in Eq.(\ref{gausscc}).

As said in the introduction, in principle we could work with any one of
the structures appearing in Eq~.(\ref{trace}). To study the dependence of
our results with the chosen structure, we show in Fig.~5, through the squares,
the form factor $g_{J/ \psi D^* D^*}^{(D^*)}(Q^2)$ obtained from the
$p_\mu p_\nu p_\alpha$ structure, using the same procedure described above. 
In this case
the QCDSR results can be fited by an exponential parametrization, dotted
line in Fig.~5, given by
\beq
g_{J/ \psi D^* D^*}^{(D^*)}(Q^2)=1.96e^{-Q^2/3.55}\;,
\label{expo}
\end{equation}
which gives, at the $D^*$ pole, $g_{J/ \psi D^* D^*}=6.11$. Although the 
parametrizations in Eqs.~(\ref{monod}) and (\ref{expo}) are rather different,
the behaviour of the curves is not so different, as can be seen in Fig.~5,
and lead to consistent values for the coupling constant.
These differences can be attributed to the uncertainties of the method.

\begin{figure}[ht] 
\begin{center}
\epsfig{file=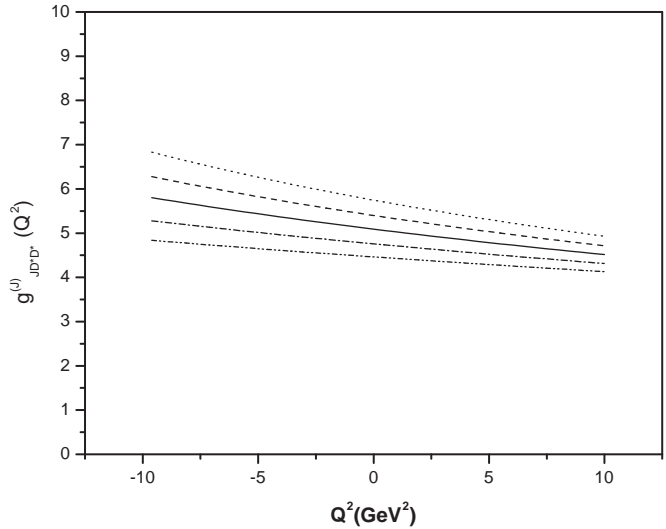} 
\caption{Dependence of the form factor with the continuum threshold for
the $J/\psi$ off-shell case. The lower curve (dash dot dot) corresponds to 
$\Delta_{s,u} = 0.5\GeV$ and the upper one (dot) to $\Delta_{s,u}= 0.7\GeV$.
The dash correspond to  $\Delta_{s,u}= 0.7,0.6\GeV$, the solid to  
$\Delta_{s,u}= 0.6,0.6\GeV$,
and the dash dot to  $\Delta_{s,u}= 0.6,0.5\GeV$}
\label{fig6}
\end{center}
\end{figure}

In order to study the dependence of our results with the continuum
threshold, we vary $\Delta_{s,u}$ between 
$0.5\GeV\le \Delta_{s,u}\le 0.7\GeV$ in the parametrization
corresponding to the $J/\psi$ off-shell case. As can be seen in 
Fig.~\ref{fig6}, this procedure gives us a uncertanty interval of 
$4.8\le g_{J/ \psi D^* D^*} \le 6.7$ for the coupling constant.

\vspace{0.5cm}

Concluding, the two cases considered here, off-shell $D^*$ or $J/\psi$,
give compatible results for the coupling constant, evaluated using the QCDSR 
approach. Considering the uncertainties in the continuum
thresholds we obtain:
\begin{equation}
g_{J/ \psi D^* D^*}= 6.2\pm0.9\;. \label{finalcoupling}
\end{equation}
 
In Table \ref{tableresults} we compare our result with those of other 
calculations. 
Our value is smaller than the values obtained using the
constituent quark-meson model \cite{deandrea} and a vector meson dominance
approach \cite{linko} and is greater than  the 
result obtained using a chiral model \cite{haglin2}. It is also
in agreement, within the errors, with the QCDSR calculation for the 
$J/\psi DD$ coupling of Ref. \cite{matheus2}, which is expected as a 
consequence of the spin symmetry of the HQET. 

\begin{table}[h]
\begin{tabular}{|c|c|c|c|c|c|}  \hline
Coupling &This work & Ref.\cite{linko} & Ref.\cite{matheus2} & 
Ref.\cite{deandrea} & Ref.\cite{haglin2}\\ \hline
$g_{J/\psi D^*D^*}$ & $6.2\pm0.9$&7.6
&$5.8\pm 0.8$&$8.0\pm 0.5$&
4.9 \\ \hline
\end{tabular}
\caption{Values of the coupling using different approaches.
Ref.\cite{linko} is a vector meson dominace calculation,
Ref.\cite{matheus2} is a QCDSR calculation in the same spirit as the
calculation presented in this paper, Ref.\cite{deandrea} uses the
constituent quark model and Ref.\cite{haglin2} uses a quiral model for
computing the coupling.}
\label{tableresults}
\end{table}

From the parametrizations in Eqs.~(\ref{gausscc}) and (\ref{monopole}) we can 
also extract the cutoff parameter, $\Lambda$, associated with the form factors.
We get $\Lambda\sim2.2\GeV$ for an off-shell $D^*$ meson and  $\Lambda\sim
8.9\GeV$ for an off-shell $J/\psi$. The cutoffs obtained here follow the same
trend as observed in Refs.~\cite{ddpir,ddrho,matheus1,matheus2}: the value of 
the
cutoff is directly associated with the mass of the off-shell meson probing the
vertex. The form factor is harder if the off-shell meson is heavier. Therefore,
in the $J/\psi D^*D^*$ vertex, as in the $J/\psi DD$, the $J/\psi$ behaves like
a point-like parton penetrating the $D^*$ meson.

\acknowledgments
This work has been supported by CNPq, CAPEs and FAPESP.



\begin{thebibliography}{99}
\bibitem{kha} D. Kharzeev, hep-ph/0408091 and references therein.
\bibitem{phenix} PHENIX Collab., S.S. Adler {\em et al.}, {\sl Phys. Rev.} 
{\bf C69}, 014901 (2004); {\sl Phys. Rev. Lett.} {\bf 92}, 051802 (2004);
R.G. de Cassagnac, nucl-ex/0403030.
\bibitem{NA50} M. A. Abreu {\em et al.} ( NA50 Collab.), 
{\sl Phys. Lett.} {\bf B477}; (NA38 Collab.), {\sl Phys. Lett.} 
{\bf B466},408 (1999).
\bibitem{mamu} S.G. Matinyan and B. M\"uller, {\sl Phys. Rev.} {\bf C58}, 
2994 (1998). 
\bibitem{linko} Z. Lin and C.M. Ko, {\sl Phys. Rev.} {\bf C62}, 034903 (2000); 
Z. Lin, C.M. Ko and B. Zhang, {\sl Phys. Rev.} {\bf C61}, 024904 (2000).
\bibitem{haglin} K.L. Haglin, {\sl Phys. Rev.} {\bf C61}, 031902 (2000).
\bibitem{su} Y. Oh, T. Song and S.H. Lee, Phys. Rev.
              {\bf C63}, 034901 (2001)
\bibitem{regina} F.S. Navarra, M. Nielsen and M.R. Robilotta,
{\sl Phys. Rev.} {\bf C64}, 021901 (R) (2001);
R.S. Azevedo, M. Nielsen, {\sl Phys. Rev.} {\bf C69}, 035201 (2004).
\bibitem{svz} M. A. Shifman, A. I. Vainshtein and V. I. Zakharov, 
{\sl Nucl. Phys.} {\bf B120}, 316 (1977).
\bibitem{ddpi} F.S. Navarra, M. Nielsen, M.E. Bracco, M. Chiapparini and
C.L. Schat, {\sl Phys. Lett.}  {\bf B489}, 319 (2000). 
\bibitem{ddpir} F. S. Navarra, M. Nielsen, M. E. Bracco, 
{\sl Phys. Rev.} {\bf D65}, 037502 (2002).
\bibitem{ddrho} M. E. Bracco, M. Chiapparini, A. Lozea, F. S. Navarra and 
M. Nielsen, {\sl Phys. Lett.} {\bf B521}, 1 (2001).
\bibitem{matheus1} R.D. Matheus, F.S. Navarra, M. Nielsen and R.R. da Silva, 
{\sl Phys. Lett.} {\bf B541}(3-4), 265 (2002).
\bibitem{matheus2} R.D. Matheus, F.S. Navarra, M. Nielsen and R.R. da
Silva, {\sl hep-ph/0310280.} 
\bibitem{dasilva} R.R. da Silva, R.D. Matheus, F.S. Navarra and M.
Nielsen, {\sl hep-ph/0310074.}
\bibitem{Angelo} A. C. da Cunha Jr, M. E. Bracco, A. Lozea, Proceedings of the 
Hadrons Physics and
Relativistic Aspects in Nuclear Physics, 2004, to appear in AIP.
\bibitem{ddpiex} CLEO Collab., S. Ahmed {\em et al.}, Phys. Rev. Lett.
{\bf87}, 251801 (2001).
\bibitem{deandrea} A. Deandrea, G. Nardulli and D. Polosa, 
hep-ph/0302273.
\bibitem{haglin2} K.L. Haglin and C. Gale, {\sl Phys. Rev.} {\bf C63}, 065201
(2001).
\bibitem{io2} B.L. Ioffe and A.V. Smilga, {\sl Nucl. Phys.} {\bf B216} 373
(1983); {\sl Phys. Lett.} {\bf B114}, 353 (1982).
\bibitem{khod} Khodjamirian {\em et al.}, {\sl Phys. Lett.} {\bf B457}, 
245 (1999). 
\end{thebibliography}
\end{document}